\documentclass[smallextended]{svjour3}       
\smartqed  
\usepackage{graphicx}
\usepackage{subfigure}
\usepackage{xurl}
\usepackage{hyperref}
\usepackage{cite}
\begin{document}

\title{Performance of Superconducting Quantum Computing Chips under Different
Architecture Designs}
\titlerunning{Performance of Quantum Computing Chips under Different Architecture Designs}        

\author{Wei Hu*         \and
        Yang Yang*\and
        Weiye Xia\and
        Jiawei Pi\and
        Enyi Huang\and
        Xin-Ding Zhang\and
        Hua Xu \thanks{ * Wei Hu and Yang Yang contributed equally to
          the work.  \\
        Corresponding author: Hua Xu
        hua.xu@kfquantum.com \and Xin-Ding Zhang}
}

\institute{Wei Hu \and Yang Yang\and Weiye Xia
  \and Hua Xu\at
              Kunfeng Quantum Technology Co., Ltd, Shanghai, China. \\
              \email{wei.hu@kfquantum.com}             \\
              \email{yang.yang@kfquantum.com}             \\
              \email{hua.xu@kfquantum.com}           
           \and
           Jiawei Pi \and Enyi Huang\and Xin-Ding Zhang
           \at
                Guangdong Provincial Key Laboratory of Quantum Engineering and Quantum
  Materials, School of Physics and Telecommunication Engineering, South China
  Normal University, Guangzhou 510006, China \\
 \email{xdzhang2000@gmail.com}
}

\date{Received: date / Accepted: date}

\maketitle

\begin{abstract}
Existing and near-term quantum computers can only perform two-qubit gates
between physically connected qubits. Research has been done on compilers to
rewrite quantum programs to match hardware constraints. However, the quantum
processor architecture, in particular the qubit connectivity and topology, still
lacks enough discussion, while it potentially has a huge impact on the
performance of the quantum algorithms. We perform a quantitative and
comprehensive study on the quantum processor performance under different qubit
connectivity and topology. We select ten representative design models with
different connectivities and topologies from quantum architecture design space
and benchmark their performance by running a set of standard quantum algorithms.
It is shown that a high-performance architecture almost always comes with a
design with a large connectivity, while the topology shows a weak influence on
the performance in our experiment. Different quantum algorithms show different
dependence on quantum chip connectivity and topologies. This work provides
quantum computing researchers with a systematic approach to evaluating their
processor design.

\keywords{Quantum Computation \and Quantum Chip Architecture \and Performance}
\PACS{07.05.Bx, 03.67.Lx}
\end{abstract}

\section{Introduction}
\label{intro}
Inspired by the vast potential applications and superb computing power, quantum
computing (QC) has attracted rapidly growing interest of researchers in the past
decades. There are multiple potential quantum computing hardware systems under
study. Among those systems, superconducting qubits
\cite{barends_superconducting_2014,corcoles_demonstration_2015,
riste_detecting_2015,ofek_extending_2016,takita_demonstration_2016}, and trapped
ions \cite{debnath_demonstration_2016,monz_realization_2016} have been leading
the technology advancement of the QC on the functionality and technology
maturity. Both systems have been able to integrate qubits on the order of tens
of qubits to nearly one hundred \cite{jurcevic_demonstration_2021-1,
  honeywell_2020,OSA_2020,
collaborators_hartree-fock_2020}, and fully programmable multi-qubit machines
have been built based on these systems. Such machines provide users with a
high-level interface that enables them to implement arbitrary quantum circuits.
This makes it possible for the first time to test quantum computers irrespective
of their particular physical implementation.

Since there are multiple physical realizations of QC and different
implementation methods of physical systems, quantum computers not only have
different number of qubits, but also different connectivity and different gating
operation between qubits. How to accurately compare and evaluate the performance
of quantum chips is then a challenge, as well as a core question. To tackle this
question, IBM proposed the concept of Quantum Volume \cite{cross_validating_2019}, and
other research groups have also come up with methodologies to evaluate and benchmark
quantum chips \cite{cross_validating_2019,paul_quantum_2020}.

It has been widely agreed that the capacity of a quantum computer is not just
determined by the number of qubits \cite{kjaergaard_superconducting_2020}. There are many
other factors, such as qubit quality including single-qubit coherent time,
single-qubit gating fidelity and two-qubit gating operation fidelity, etc, and
chip architecture including qubit-qubit connectivity, and qubit-qubit topology
layout. In addition to the number of qubits, the researchers have been mostly
focused on the quality of those qubits when discussing the performance of
quantum computers.

Recently, there are only few studies on the quantum chip architecture,
especially \textit{connectivity} and \textit{topology}. The quantified
definition of connectivity will be given in Sec. \ref{arc} while the word
topology here refers to the topological property of qubits network and has
nothing to do with topological quantum computing. Norbert M. Linke et al.  have
compared two quantum computing architectures, superconducting transmon system
and ion trap system. In their research, they pioneered the impact of
connectivity on the quantum processor performance, and claimed achieving high
connectivity for a large-scale superconducting processor is an important, but
still open question \cite{linke_experimental_2017}.  On a partially connected architecture, the
compiler must dynamically map logical qubits in an arbitrary quantum circuit to
physical qubits on the quantum chip. This problem is known as qubit allocation
or qubit mapping, which multiple methods have been developed by researchers
\cite{siraichi_qubit_2018-1,zulehner_efficient_2018,
li_tackling_2019}.

Till now, there still lacks systematic and comprehensive study on the
architecture of the quantum processor and its related connectivity and topology
to answer the following questions: how connectivity strength and position, as
well as the topology will impact the performance of a quantum processor? What
will the overhead of a certain connectivity quantum processor be?

In this paper, we have done a quantitative and comprehensive study on these
questions. We studied the design of quantum chip architecture, including
qubit-qubit connectivity and qubit topology layout, and analyzed the performance
of different architecture designs both qualitatively and quantitatively. Our
study will be helpful to give quantitative analysis and comparison of different
QC systems. Besides, this study will dramatically help the QC researchers to
design their chips. With a systematic method to evaluate the chip performance,
the QC researchers can balance the connectivity requirement and other
restrictions when designing the QC chips, to achieve the best performance in
their design space.

Whereas the quantum computers considered here are still small scale and their
capabilities do not currently reach beyond demonstrative algorithms, this line
of inquiry can still provide useful insights into the performance of existing
systems and the role of architecture in quantum computer design. These findings will be
crucial for the realization of more advanced future incarnations of the present
technologies.

The organization of the paper is as follows:

1. Describe the overall experimental setup. It includes the workflow of the
 experiment, and how we design the quantum algorithm benchmark suite for the
evaluation.

2. Based on the results of different chip architectures, we perform the
analysis and evaluate the impact of qubit connectivity on
the performance of the benchmarks.

3. Finally, based on the analysis, we propose some guidelines on the quantum
processor design. Moreover, some possible future topics have also been
proposed. The method developed in the paper will be a great contribution for
QC designers.

\section{Experiment} \label{experiment}

\begin{figure*}[htpb]
  \centering
  \includegraphics[width=0.8\linewidth]{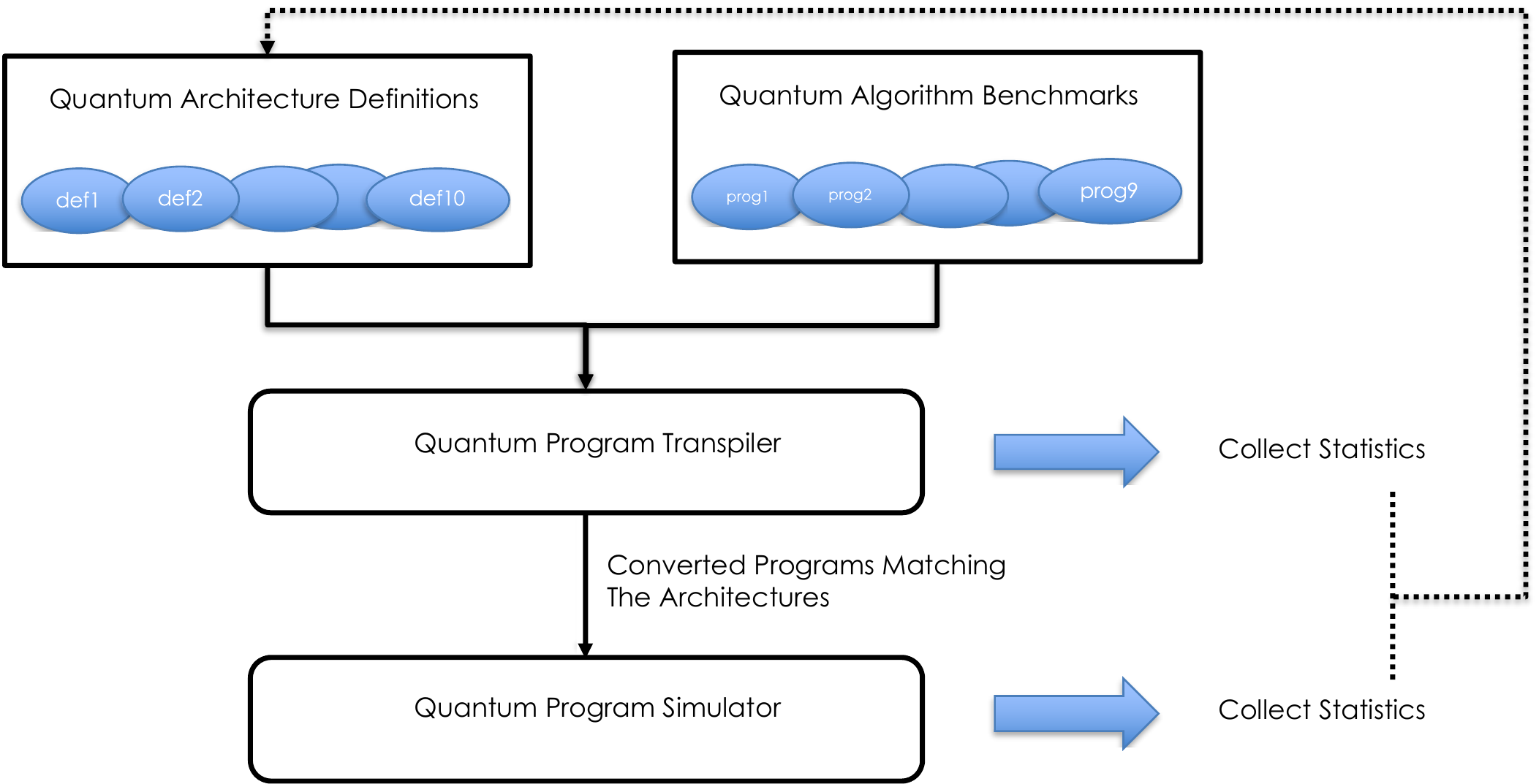}
  \caption{Experiment setup and overall workflow.}%
  \label{fig:expfig1}
\end{figure*}

The workflow of our experiment is depicted in Fig. \ref{fig:expfig1}.
To study the performance of different quantum architectures, we evaluate 10
representative architectures based on existing designs, by running 9 well-known
quantum algorithm implementations as benchmarks. We make use of IBM Qiskit to transpile
and simulate the benchmarks on every architecture. The quantitative analysis of the performance
metrics collected offers valuable guidance to quantum chip designers.

\subsection{Selection of Algorithms}

Inspired by a previous work \cite{li_tackling_2019}, we select $5$ representative quantum
algorithms for testing, shown in Table.
\ref{tab:1}. Most of these algorithms are commonly viewed as classical quantum
algorithms, including Quantum Fourier Transfrom (QFT), Quantum Phase Estimation
(QPE), Surface Code Error Correction (SCE).  Additionally, the problem of one
dimensional Ising model with six qubits is chosen as another test algorithm.

The QFT and QPE are the cornerstones of many other algorithms, such as Shor's
algorithm, Quantum Machine Learning algorithms and Quantum option pricing
algorithms \cite{nielsen_quantum_2010-1, lloyd_quantum_2013-1,
stamatopoulos_option_2020-1, ramos-calderer_quantum_2021-1}. Quantum error
correction (QEC) is a central topic in quantum information theory so we choose
two test algorithms, namely QEC with Steane-enlargement and the surface code
\cite{christensen_steane-enlargement_2020-1, fowler_surface_2012-1}. The surface code is used for fault-tolerant quantum
computation. The code requires a 2D square-lattice of qubits with only the
nearest neighbor interactions.  The Ising model is one of the most studied
models in statistical physics, like the hydrogen atom model in quantum mechanics
\cite{brush_history_1967-1}.

For the algorithms we have selected, we create their implementations including $qft_{12}$,
$qft_{16}$, $qft_{30}$, $qft_{32}$, $qpe_{15}$, $steane_{25}$, $surface_{15}$,
$surface_{25}$, $ising_6$, where the subscript indicates the number of qubits used
and the meaning of abbreviations is summarized in Table. \ref{tab:1}.

\begin{table*}[htpb]
  \centering
  \begin{tabular}{|c|l|}
  \hline
  abbreviations & \multicolumn{1}{c|}{algorithms} \\ \hline
  $qft$         & quantum fourier transform       \\ \hline
  $qpe$         & quantum phase estimation        \\ \hline
  $steane$      & quantum error-correcting codes obtained by using Steane-enlargement                              \\ \hline
  $surface$     & quantum error correction with the surface code                    \\ \hline
  $ising$       & one dimensional Ising model     \\ \hline
  \end{tabular}
  \caption{Abbreviations.}
  \label{tab:1}
\end{table*}

\subsection{Selection of Architectures}
\label{arc}

There are several candidate technologies to implement QC physically, including
superconducting quantum circuit \cite{clarke_superconducting_2008-1}, ion trap
\cite{kielpinski_architecture_2002-1, cirac_quantum_1995-1}, quantum dot
\cite{imamoglu_quantum_1999-1}, neutral atom \cite{henriet_quantum_2020-1,
igeta_quantum_1988-1}, etc. Among them, the superconducting quantum circuit is
the most promising one with IBM and Google as two leading players in this field.
IBM has built the first commercialized quantum computing platform since 2015.
Google announced a 72-qubit chip in 2018 and claimed quantum supremacy on a
53-qubit quantum processor in 2019 \cite{arute_quantum_2019-1}. There are two
well-known types of connectivity in quantum chip design, Low Connectivity (LC)
and Linear Nearest Neighbor (LNN). The former is adopted by IBM's Almaden,
Boeblinden, Singapore, Johannesburg, Poughkeepsie and Tokyo architectures, while
the latter, another popular design, is used by Google in their 53- and 72-qubit
systems \cite{arute_quantum_2019-1}. Google Sycamore chooses a square-like
structure and Google Bristlecone is a rectangle-like $6\times12$ lattice
structure.

\begin{figure*}[htpb]
  \centering
  \subfigure[\ $r_1$ architecture]{
    \includegraphics[width=0.32\linewidth]{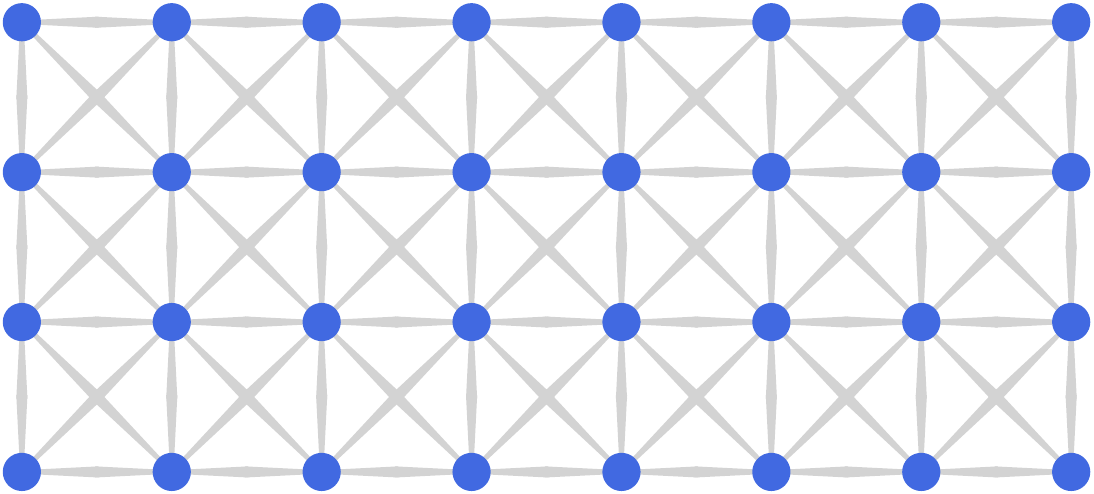}}
  \subfigure[\ $r_2$ architecture]{
    \includegraphics[width=0.32\linewidth]{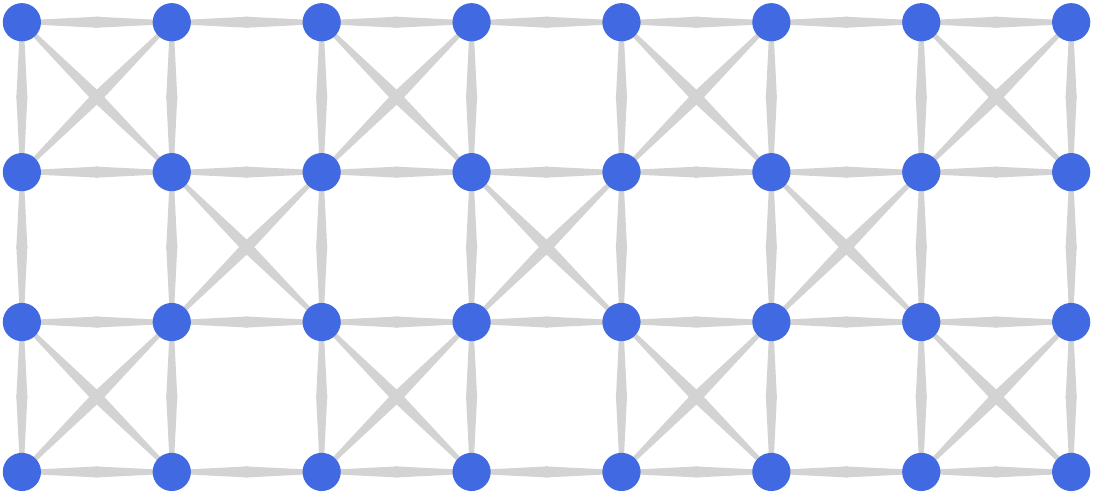}}
  \subfigure[\ $r_3$ architecture]{
    \includegraphics[width=0.32\linewidth]{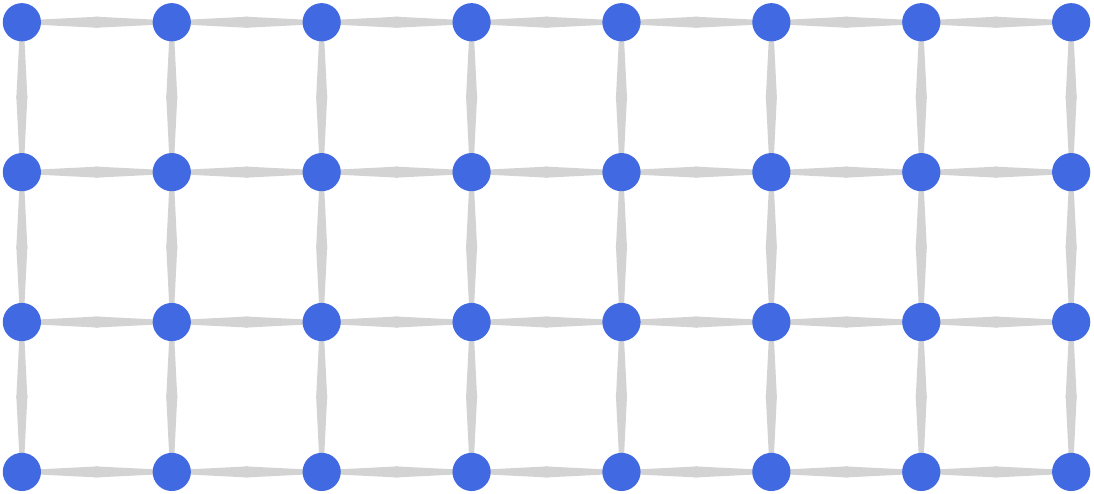}}
  \subfigure[\ $r_4$ architecture]{
    \includegraphics[width=0.32\linewidth]{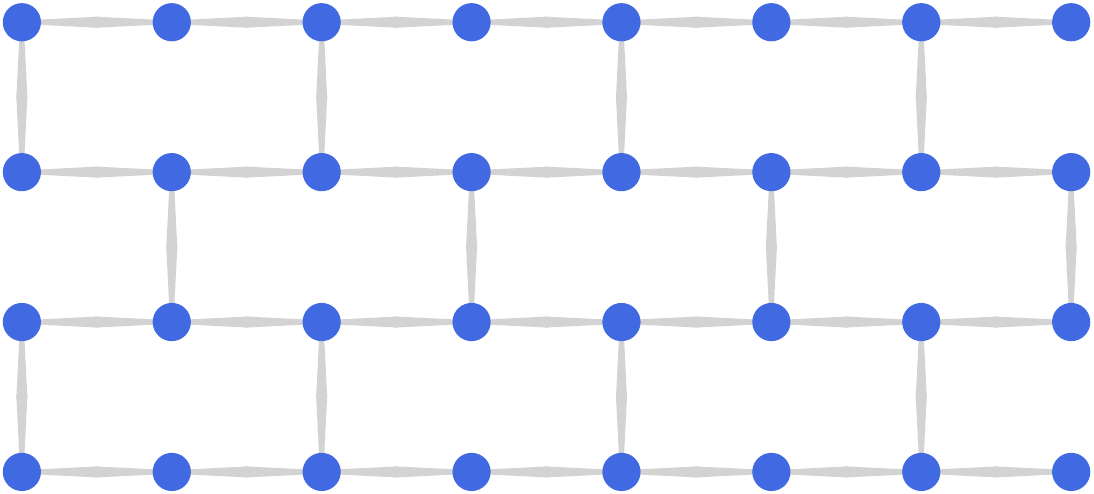}}
  \subfigure[\ $r_5$ architecture]{
    \includegraphics[width=0.32\linewidth]{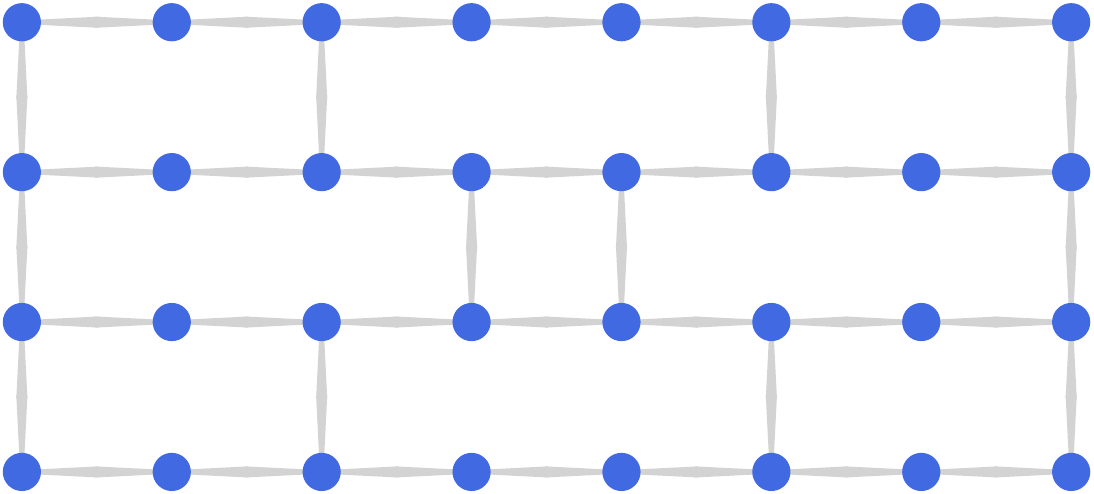}}
  \vfill
  \subfigure[\ $s_1$ architecture]{
    \includegraphics[width=0.3\linewidth]{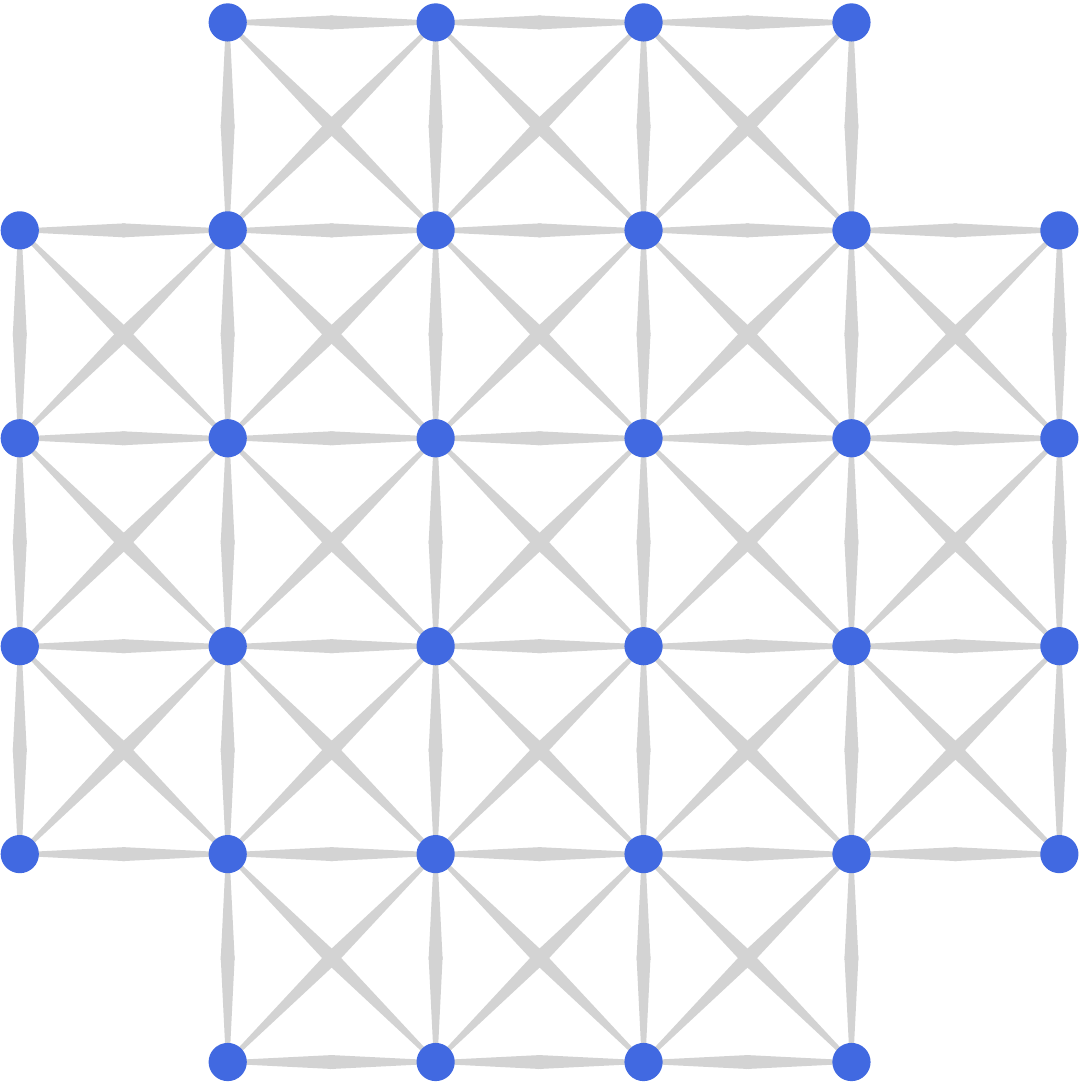}}
  \subfigure[\ $s_2$ architecture]{
    \includegraphics[width=0.3\linewidth]{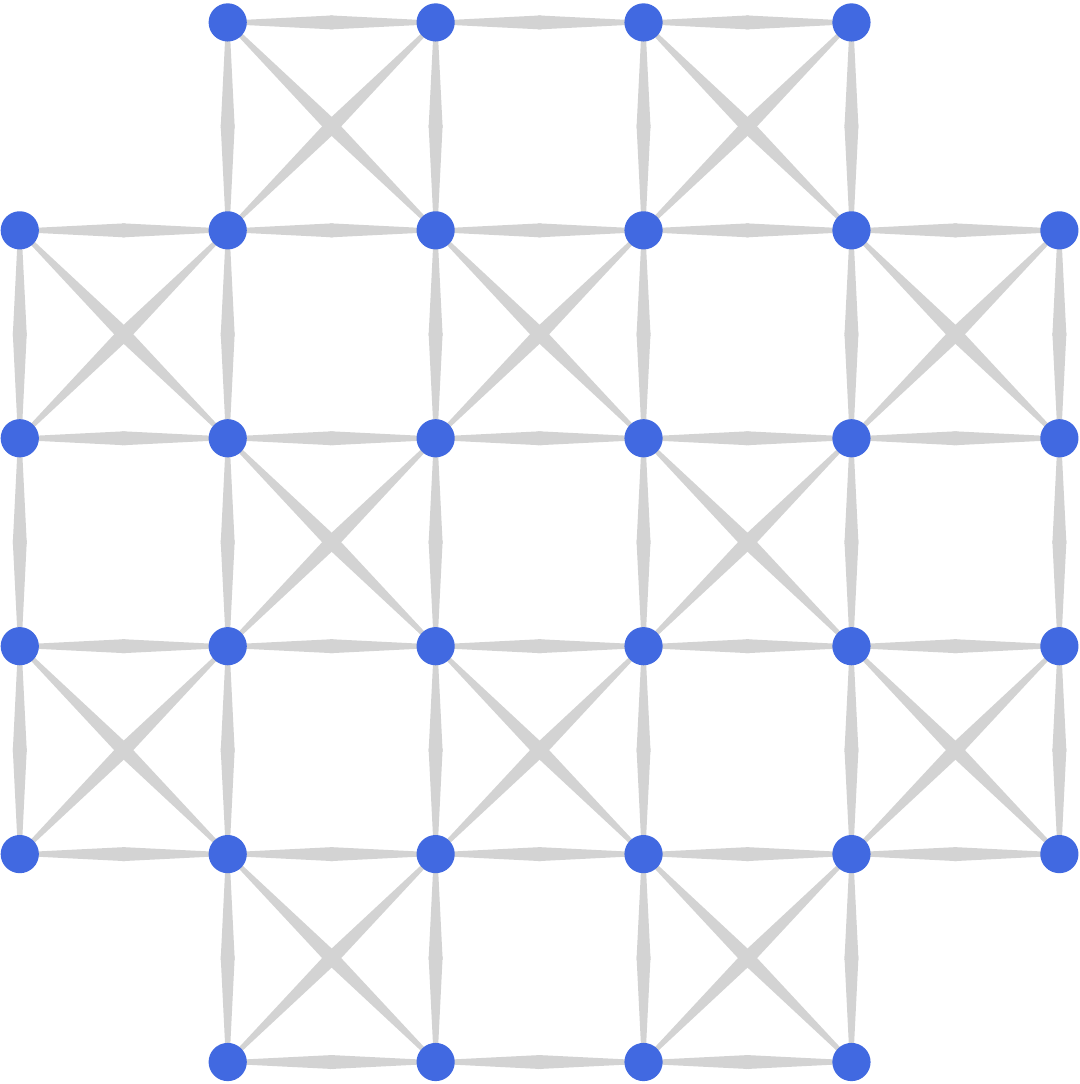}}
  \subfigure[\ $s_3$ architecture]{
    \includegraphics[width=0.3\linewidth]{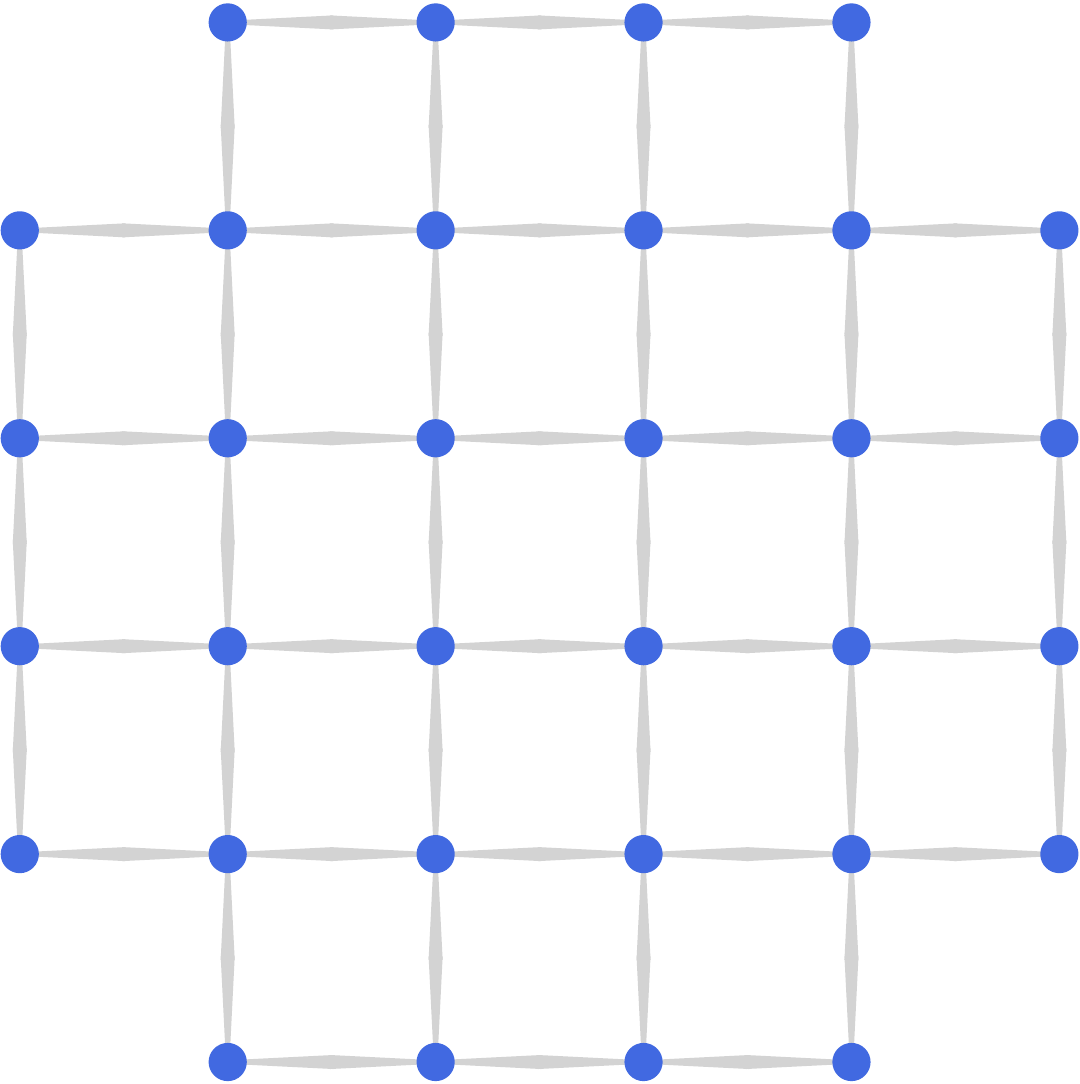}}
  \subfigure[\ $s_4$ architecture]{
    \includegraphics[width=0.3\linewidth]{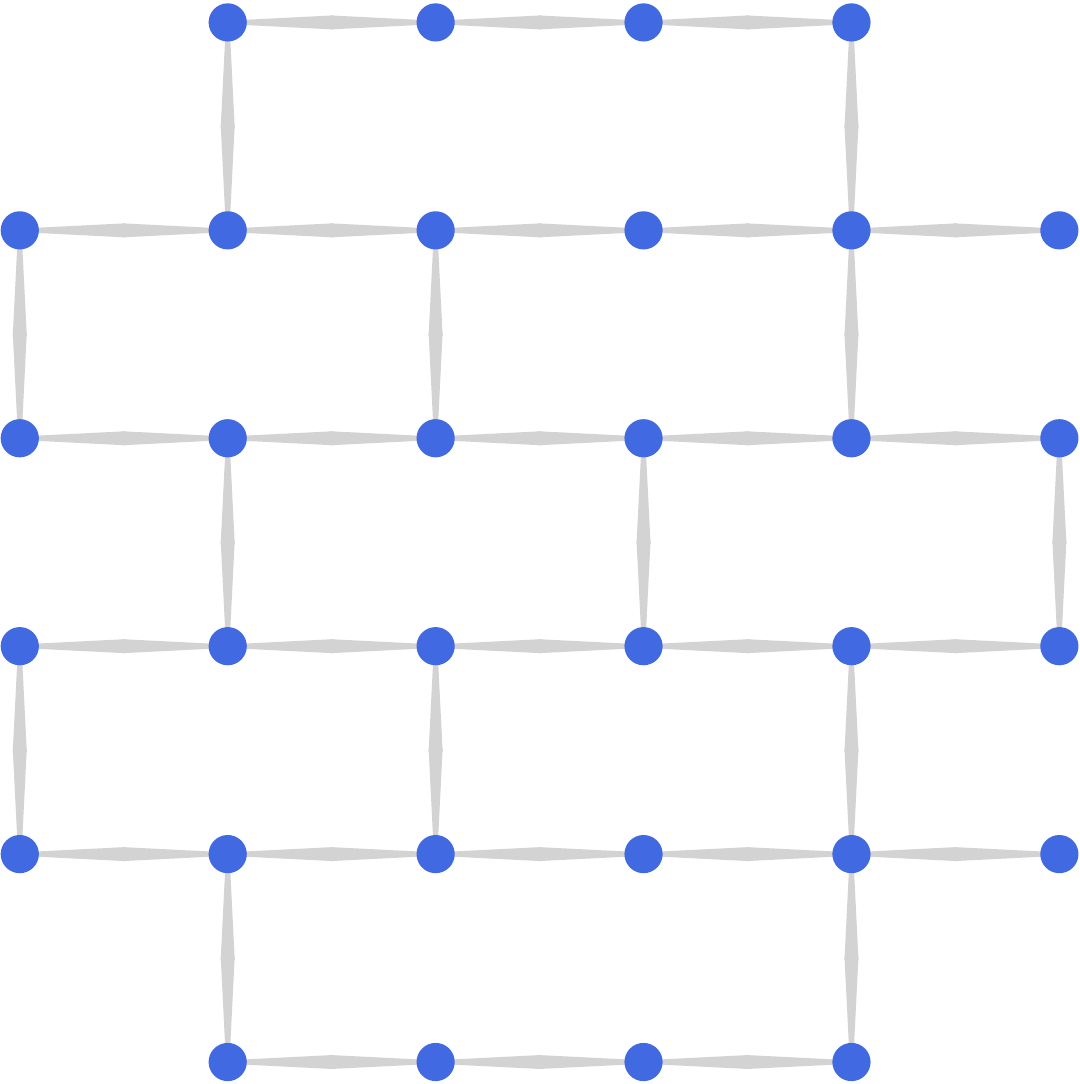}}
  \subfigure[\ $s_5$ architecture]{
    \includegraphics[width=0.3\linewidth]{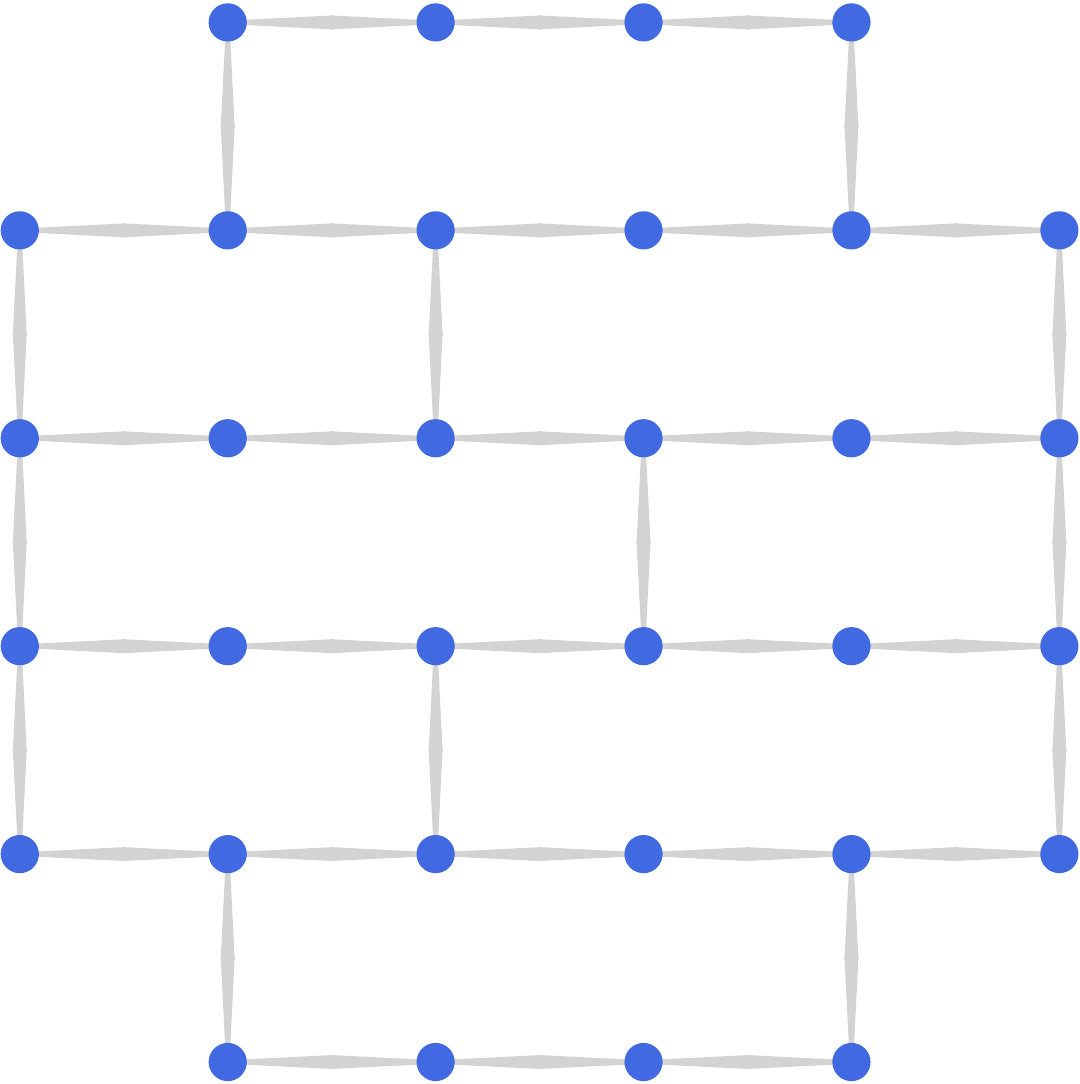}}
  \caption{The schematics of five rectangle-like and five  square-like
  architectures.}%
  \label{fig:s-and-r}
\end{figure*}

Inspired by these pioneer designs, we propose two types of architecture models
of 32 qubits that meet physics constraints, shown in Fig.  \ref{fig:s-and-r}.
They demonstrate the differences in topology and connectivity.  The first five
chips labeled $r_1$ through $r_5$ have a rectangle-like topology, whereas the
other five labeled $s_1$ through $s_5$ have a square-like topology. $r_1$ is a
fully connected variant based on LNN.  $r_2$ is the IBM Q 20 Tokyo architecture.
$r_3$ is the commonly used LNN, adopted by Google Sycamore and Bristlecone.
$r_4$ and $r_5$ are similar to the architecture of IBM Almaden, Boeblinden,
Singapore, Johannesburg and Poughkeepsie. The connectivity of $r_4$ and $r_5$
are the same but the positions of connections are slightly different.  These
five architectures give four different connectivities. $s_1$ through $s_5$ are
the square-like counterparts of $r_1$ through $r_5$. The square topology is a
more symmetrical layout with a length/width aspect ratio close to $1$.  Compared
with the square-like topology, the rectangular-like topology is a more
asymmetrical layout with different aspect ratios. For instance, the aspect ratio
is $32$ for $32\times 1$ lattice topology (single-chain topology), and the
aspect ratio is $8$ for $16\times 2$ lattice topology. In this article, the
rectangular-like topology we discussed has aspect ratio of $2$ with $8\times 4$
lattice topology.

Since there are the same numbers of qubits (vertices) in the rectangle- and
square-like circuit, connectivity $c$ can be quantified as follows,
\begin{equation} c = \frac{n_{con}}{n_{full}}, \end{equation} where $n_{con}$
and $n_{full}$ indicate the number of connected edges and the number of edges in
the fully connected architecture, respectively, in these two corresponding
configurations. The values of connectivity can be found in Table \ref{tab:2}. It
is noticeable that $c_{r_1} > c_{r_2} > c_{r_3} > c_{r_4} = c_{r_5}$, and
$c_{s_1} > c_{s_2} > c_{s_3} > c_{s_4} = c_{s_5}$, and it is possible that the
corresponding performance would follow a similar relation. Our experiment will
be able to validate this intuition.  We will also try to answer other
non-intuitive questions, such as how the rectangle-like topology compares to the
square-like topology, which of $r_4$ or $r_5$ is better, which of $s_4$ or $s_5$
is better, etc.

\begin{table}[htpb]
  \centering
  \begin{tabular}{|c|c|c|c|c|c|c|c|c|c|c|}
  \hline
                             & $r_1$  & $r_2$   & $r_3$   & $r_4$  & $r_5$  & $s_1$   & $s_2$   & $s_3$   & $s_4$    & $s_5$  \\ \hline
   $n_{con}$  & $188$  & $148$   & $104$   & $80$   & $80$   &  $188$  & $152$   & $104$   & $78$     & $78$   \\ \hline
   $c$           & $1.0$ & $0.79$  & $0.55$   & $0.43$ & $0.43$ &  $1.0$  & $0.81$  & $0.55$  & $0.41$   & $0.41$ \\ \hline
  \end{tabular}
  \caption{The connectivity in the rectangle-like and square-like
    architectures. $n_{con}$ indicates No. of connected edges and
  $c$ represents the connectivity.}
  \label{tab:2}
\end{table}

In summary, we have two types of architecture topologies and for each topology,
there are five levels of connectivities. These settings are essential to reveal
how topology and connectivity impact the performance of QC processors.

\subsection{Running the Experiment}

To evaluate the performance of each architecture, we make use of the IBM Qiskit
transpiler \cite{qiskit-transpiler} that rewrites quantum programs to match the
architecture's qubit connectivity and its native quantum gates. We then run the
IBM Qiskit simulator on the original benchmark programs as well as the output
programs. We collect data as performance metrics, which will be discussed in
Sec. \ref{result-and-discussion}.

To draw a fair comparison between architectures, ideally the transpiler should find
the optimal program equivalent to the input. However, since this problem is
NP-complete \cite{siraichi_qubit_2018-1}, finding the optimal result is not always feasible.
Here we first summarize the transpilation process, and then explain our criteria of selecting
the output program that best represents the performance of a benchmark on an architecture.

A quantum program usually assumes all-to-all qubit connectivity, and freely uses
any quantum gates allowed by the programming language. In reality, a target
quantum chip only supports a handful of quantum gates, and can only perform
two-qubit gates between physically connected qubits. Barenco et al.  proved that
an arbitrary quantum gate can be decomposed into single-qubit gates and
two-qubit CNOT gates, or any set of so-called universal quantum gates
\cite{barenco_elementary_1995-1}. To work around qubit connectivity limitations, the transpiler
searches for and inserts SWAP operations when necessary till an efficient
mapping from logical qubits to physical qubits on the quantum chip are
achieved.

Qiskit ships two qubit mapping algorithms, the original, default one, and a
newer one called SABRE, proposed by Li et al \cite{li_tackling_2019}. In addition to transformations
for the target architecture, Qiskit also has transformations aimed at optimizing
the performance. Similar to classical compilers, Qiskit
comes with four optimization levels from $0$ to $3$, although more optimizations
do not always give better performance.

With $2$ choices of qubit mapping algorithm and $4$ levels of optimization,
Qiskit provides $8$ combinations. For every pair of benchmark program and target
architecture, we run each of the $8$ mapping/optimization combinations $10$
times to produce $80$ output programs in total, and then select the best output
according to the scoring function below.  The reason we run each combination
multiple times is that the output can differ over repeated runs, caused by
randomization introduced to find approximate solutions to the NP-complete qubit
mapping problem.

As the target architecture is only hypothetical with no real hardware, we
propose a scoring function that estimates the error rate of a program.  Because
noisy quantum gates introduce errors, it is important to minimize the number of
gates to reduce the accumulated errors. It is also important to minimize the
circuit depth so that all the computation can be accomplished before the qubits
lose their quantum states. This reasoning leads to the following scoring
function $s$ that estimates the overall error rate of a quantum circuit,
\begin{equation}
  s = \beta \cdot \big[1 - (1 - \frac{\sum E_{r_i} N_i}{\sum  N_i})^{depth} \big],
  \label{eq:s_full}
\end{equation}
in which the term on the right-hand side corresponds to error rates contributed
by circuit depth. $N_i$ is the number of a certain gate, and $E_{r_{i}}$ is the
corresponding error rate of this gate. $\beta$ is a coefficient determined by
data from real chips. $\frac{\sum E_{r_i} \cdot N_i}{\sum N_i}$ is the average
error rates, replacing the error rate of each individual layer for simplicity.
In this paper, given that $\frac{\sum E_{r_i} \cdot N_i}{\sum N_i} \ll 1$ and
drop the constant $\beta$, a simplified formula of Eq.  \ref{eq:s_full}
is used to estimate the overall error rates of a quantum circuit, which is
written as the multiplication of circuit depth and average error rate of each
circuit layer,
\begin{equation}
  s = depth\cdot\frac{\sum E_{r_i} \cdot N_i}{\sum N_i}.
  \label{eq:s_simplified}
\end{equation}
The Eq. \ref{eq:s_simplified} above has been adopted in our study with IBM's
error rate data, namely the single-qubit gate average error rate $3.8 \times
10^{-4}$ and two-qubit gate error rate $6.4\times 10^{-3} $
\cite{jurcevic_demonstration_2021-1}. A lower score is preferred as it implies
smaller overall error rates and better fidelity of a quantum circuit. And we use
the score to select the output program (quantum circuit) that represents the
performance of a benchmark on a given architecture.


\section{Result Analysis and discussion} \label{result-and-discussion}

In this section, we evaluate the impact of qubit connectivity and topology on
the performance. By defining the native gates on all target architectures
identical to the typical IBM Q devices, this work ignores the impact of native
gates. Through the steps described in Sec.  \ref{experiment}, we run the Qiskit
transpiler on a set of benchmark programs for ten representative architectures,
select the best output, and then run the Qiskit simulator on them. In this
experiment we collect the transpilation time $t_{trans}$, the normalized number
of gates $\hat{n}_{gate}$, the normalized depth of the circuit $\hat{d}$, and
the normalized simulation run time $\hat{t} _{sim}$. To compare the
architecture-conforming program with the original program, we normalize the data
by dividing the number after transpilation by the number of the original
version.

All experiments are executed on an AWS EC2 x1.32xlarge instance with 4 Intel
Xeon E7 8880 Haswell CPUs (128 logical cores) and 1,952GB of memory. The
Operating System is AMI Linux 2 with Linux kernel version of 4.14. The Qiskit
version is 0.20.1.

Intuitively, the four types of data we collect are all affected by connectivity,
directly or indirectly. With reduced connectivity, the transpiler needs to work
harder, thus increasing the transpilation time. When connectivity is reduced,
more gates have to be inserted to swap the qubits, which also causes the
increase in the circuit depth and the simulation time. The influence of topology
may be less intuitive.

\begin{figure*}[htpb]
  \centering
  \subfigure[\ The number of gates vs. connectivity]{
    \includegraphics[width=0.485\linewidth]{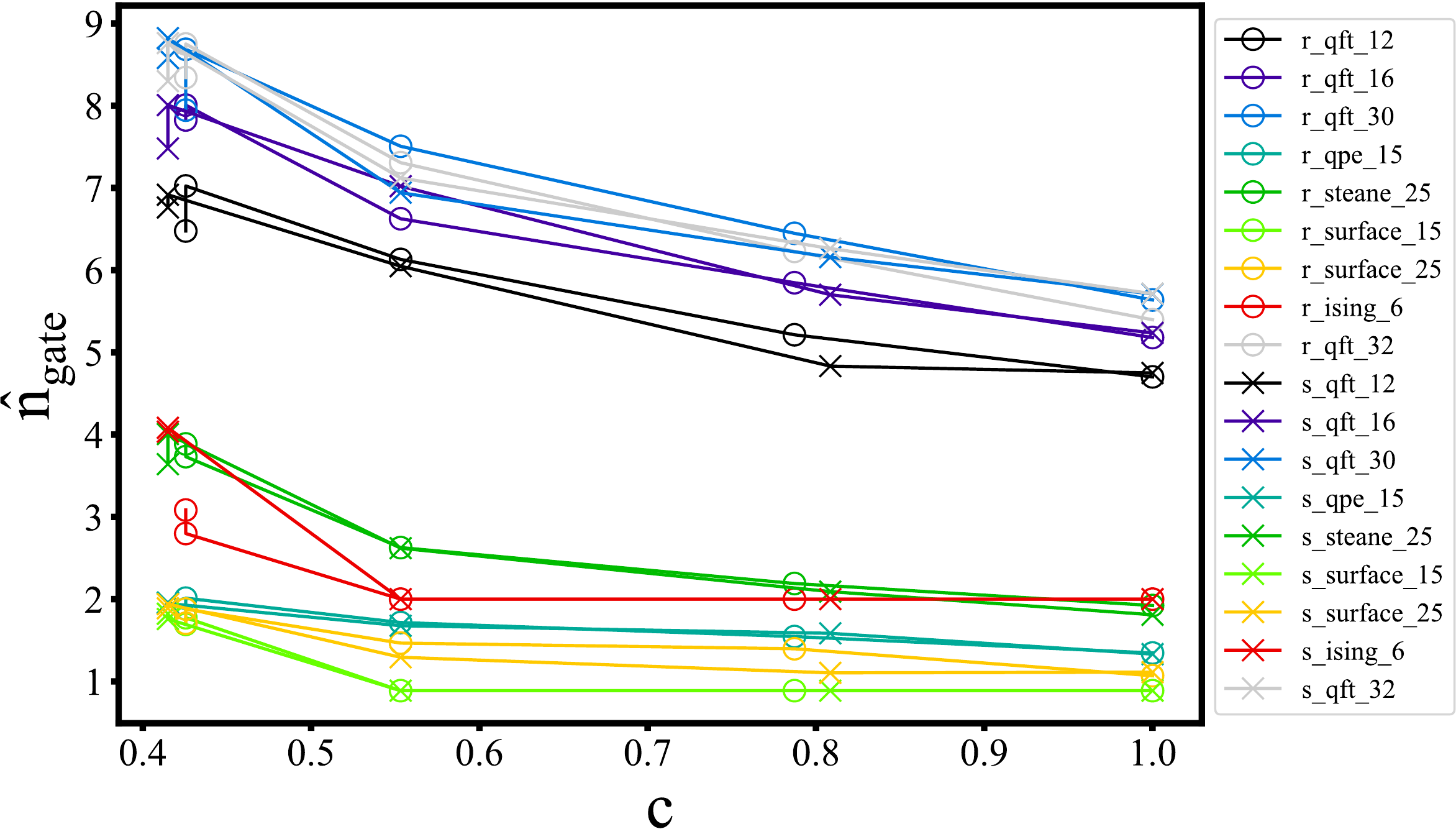}}
  \subfigure[\ The depth of circuits vs. connectivity]{
    \includegraphics[width=0.485\linewidth]{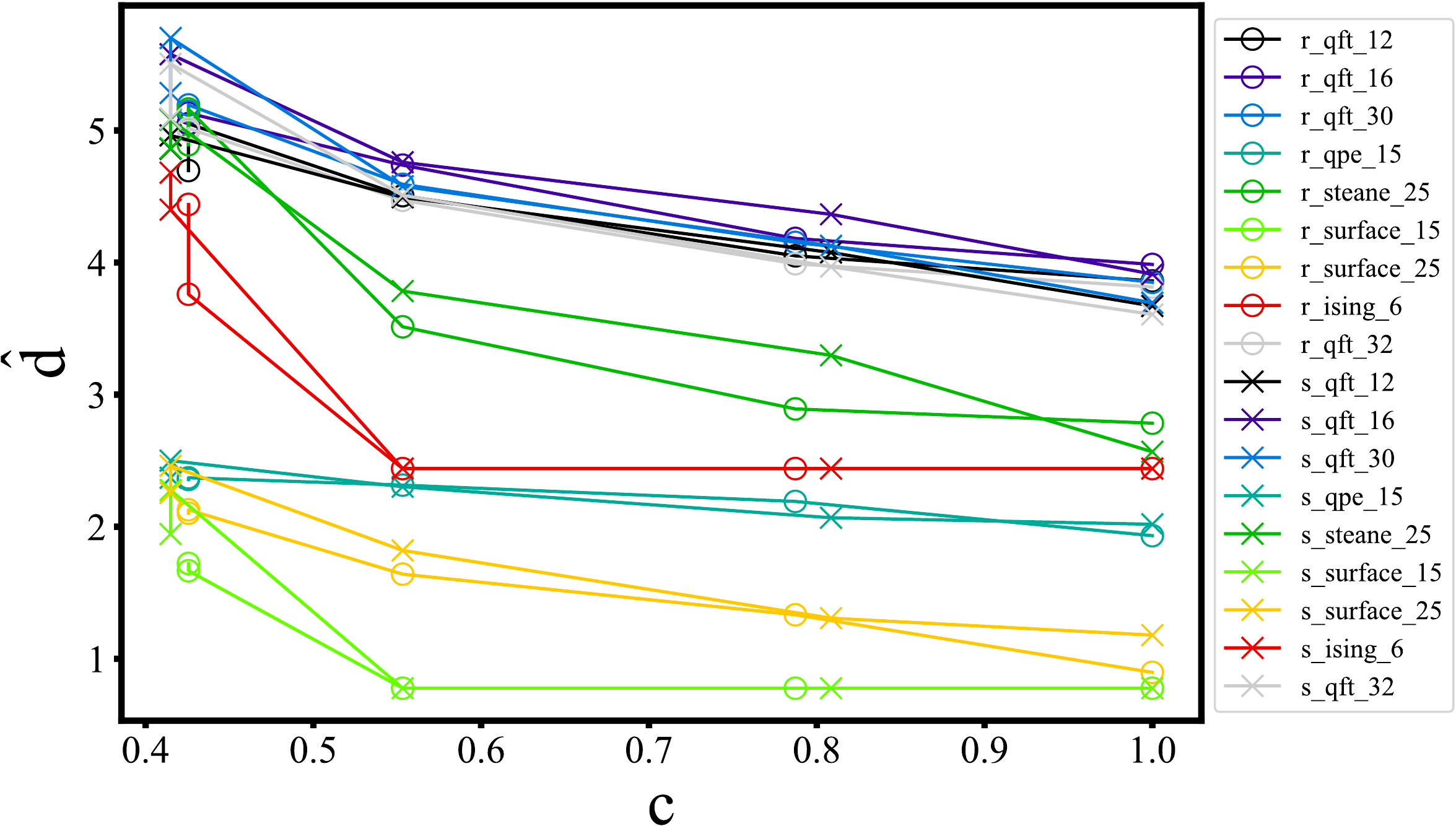}}
  \subfigure[\ The normalized score functioin vs. connectivity]{
    \includegraphics[width=0.485\linewidth]{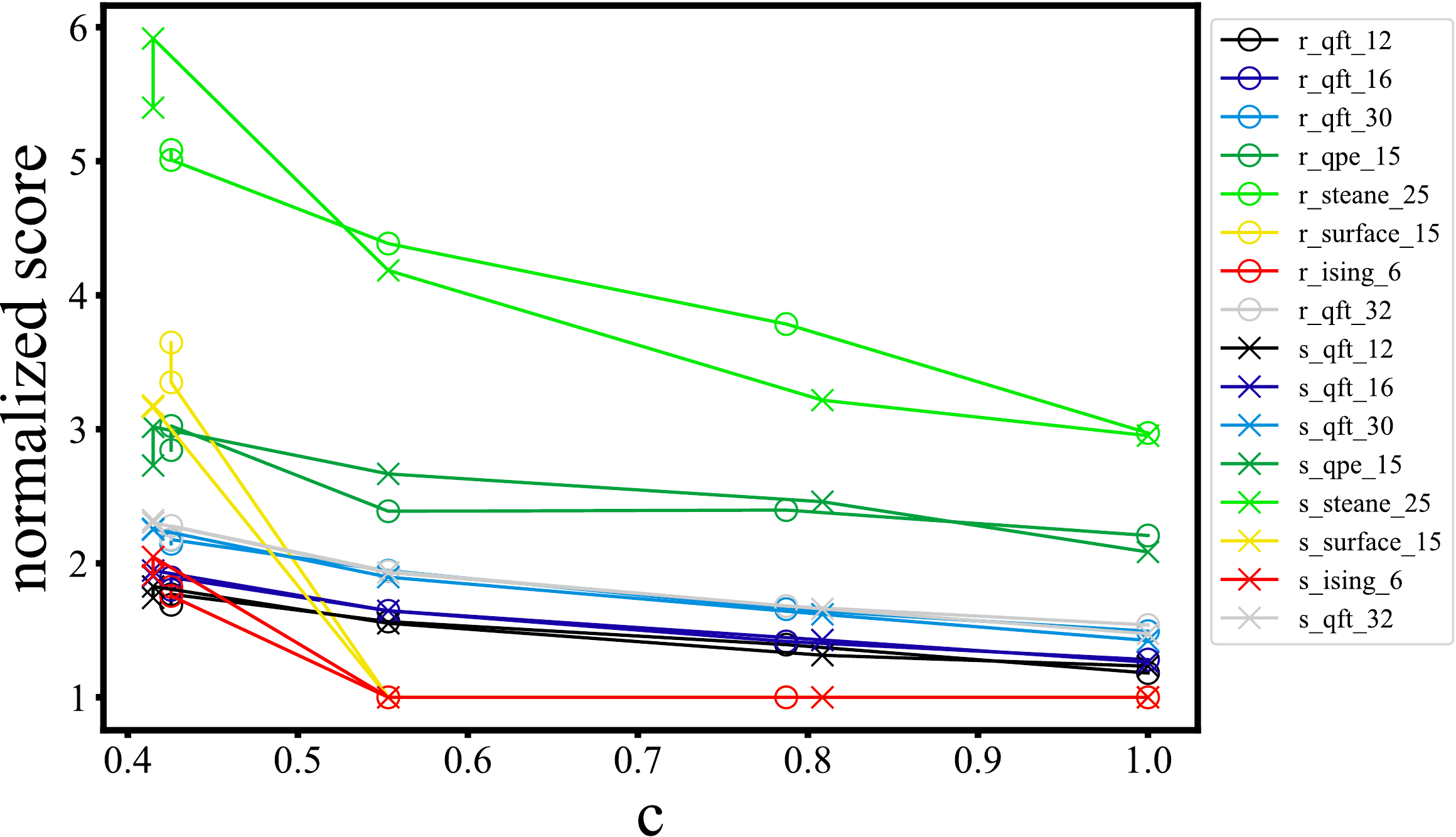}}
  \subfigure[\ The transpilation time vs. connectivity]{
    \includegraphics[width=0.485\linewidth]{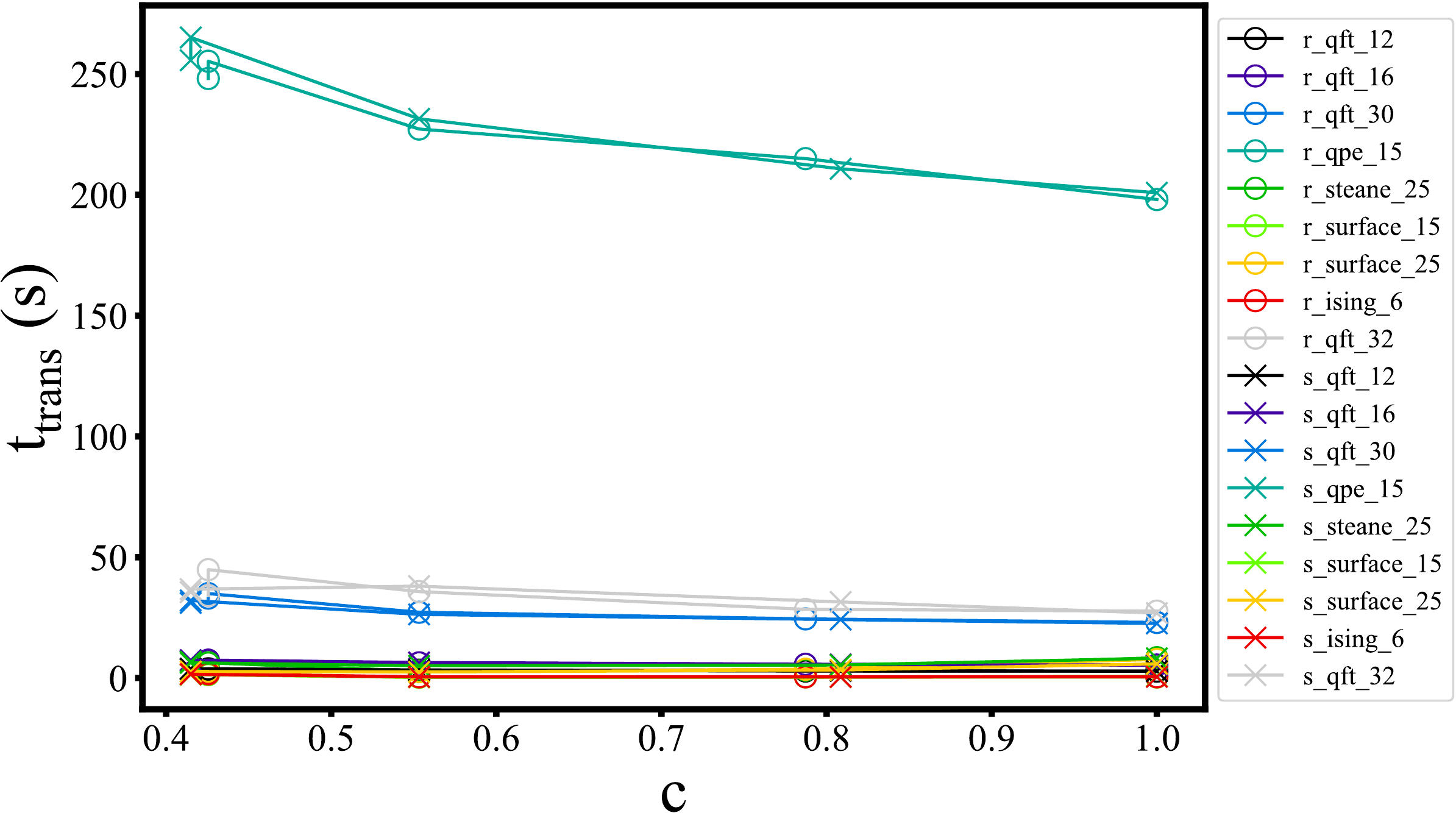}}
  \caption{
    (a) -- (d) shows the experiment results of how $\hat{n}_{gate}$, $\hat{d}$,
    normalized score functions, and $t_{trans}$ change with different connectivites $c$,
    respectively, by benchmarking the algorithms under different architectures
    ($r$ and $s$ are short for rectangle-like and square-like respectively). The
    circle markers indicate the $r$ architectures and the cross markers indicate
    the $s$ architectures. Different algorithms are labeled with different
    colors but the same algorithm under different architectures is labeled with
    the same color.
  }%
  \label{fig:data_normal_scale}
\end{figure*}

\begin{figure*}[htpb]
  \centering
  \subfigure[\ The simulation run time vs. connectivity]{
    \includegraphics[width=0.485\linewidth]{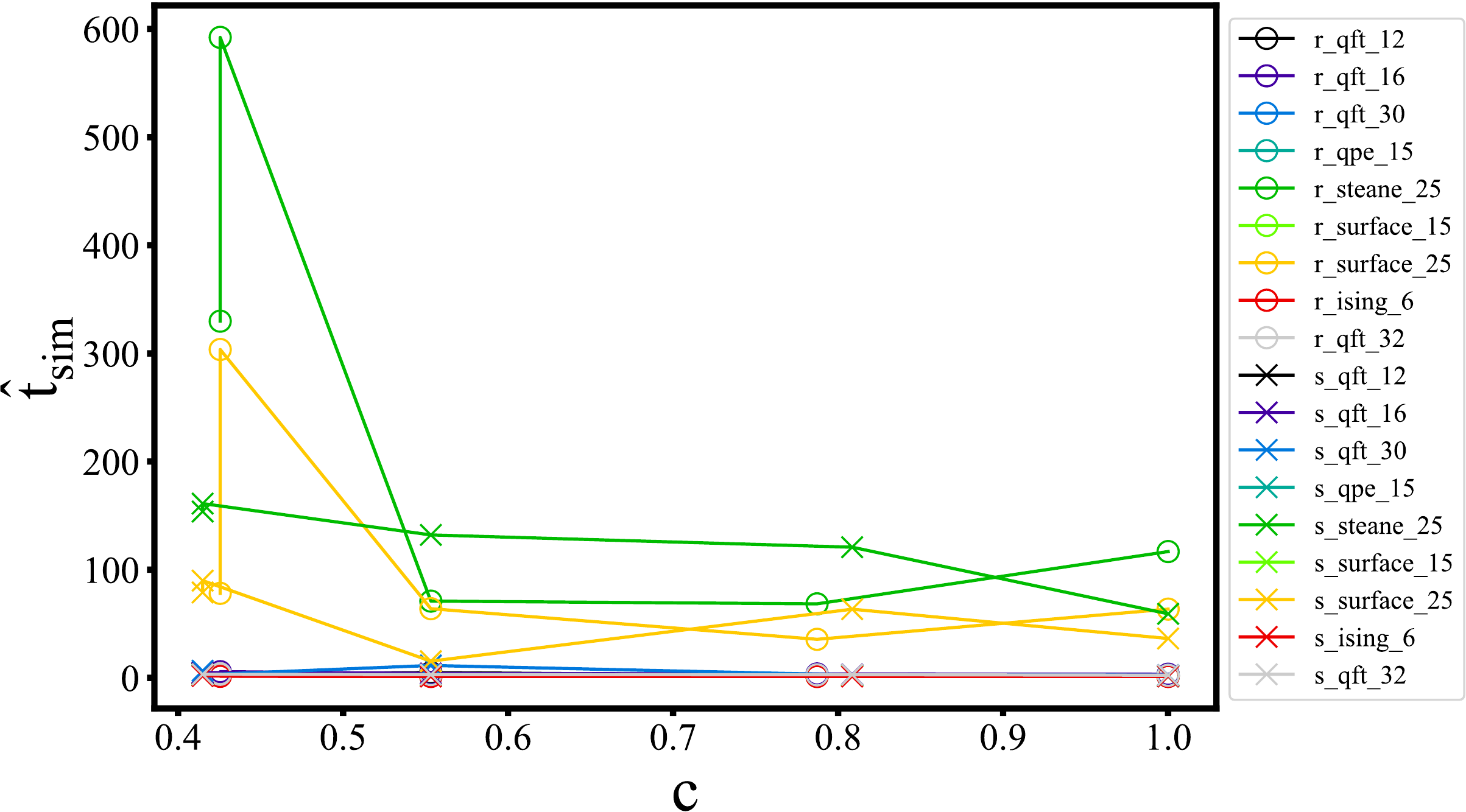}}
  \subfigure[\ The transpilation time vs. connectivity]{
    \includegraphics[width=0.485\linewidth]{fig/trans.pdf}}
  \subfigure[\ The simulation run time vs. connectivity (logy-scale)]{
    \includegraphics[width=0.485\linewidth]{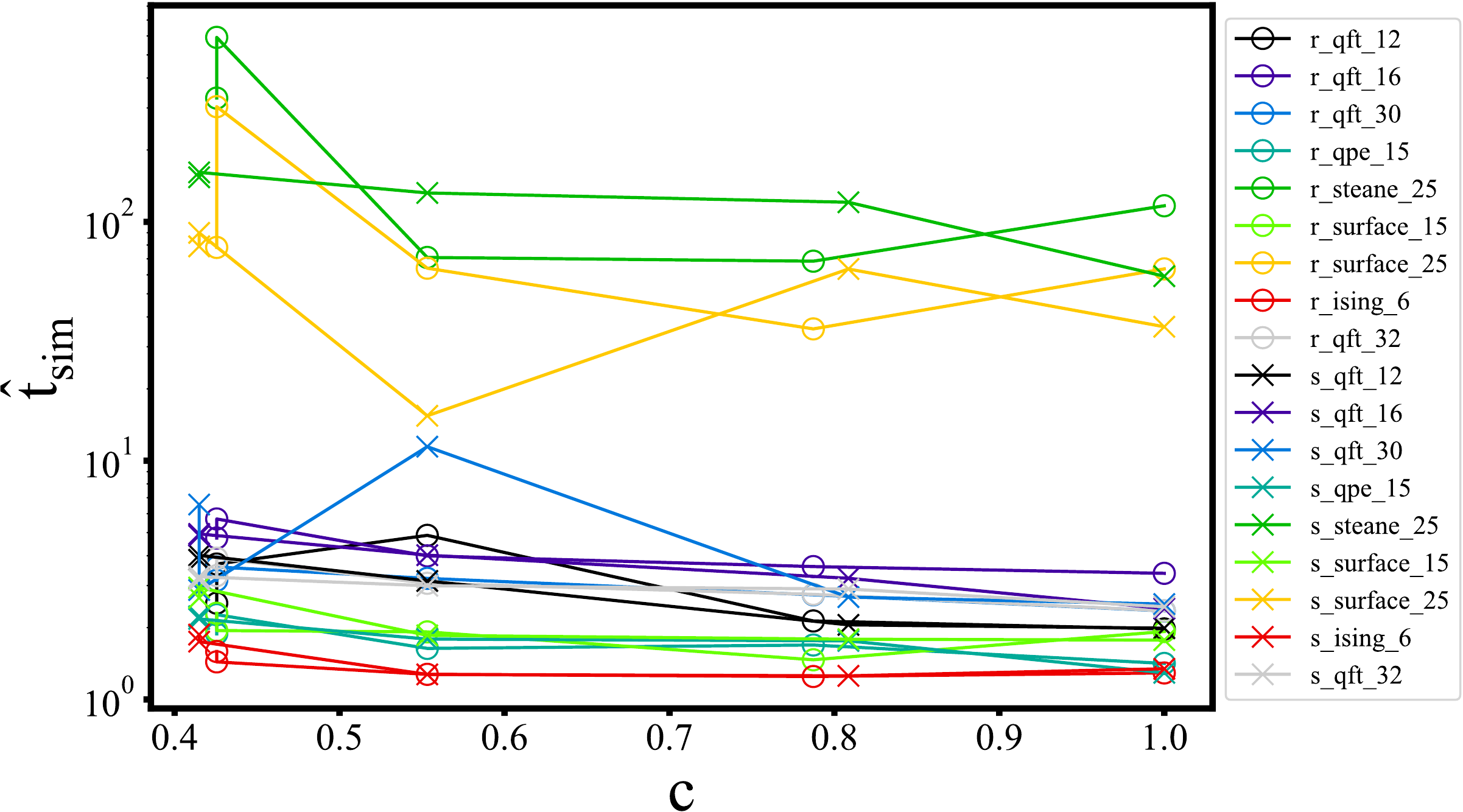}}
  \subfigure[\ The transpilation time vs. connectivity (logy-scale)]{
    \includegraphics[width=0.485\linewidth]{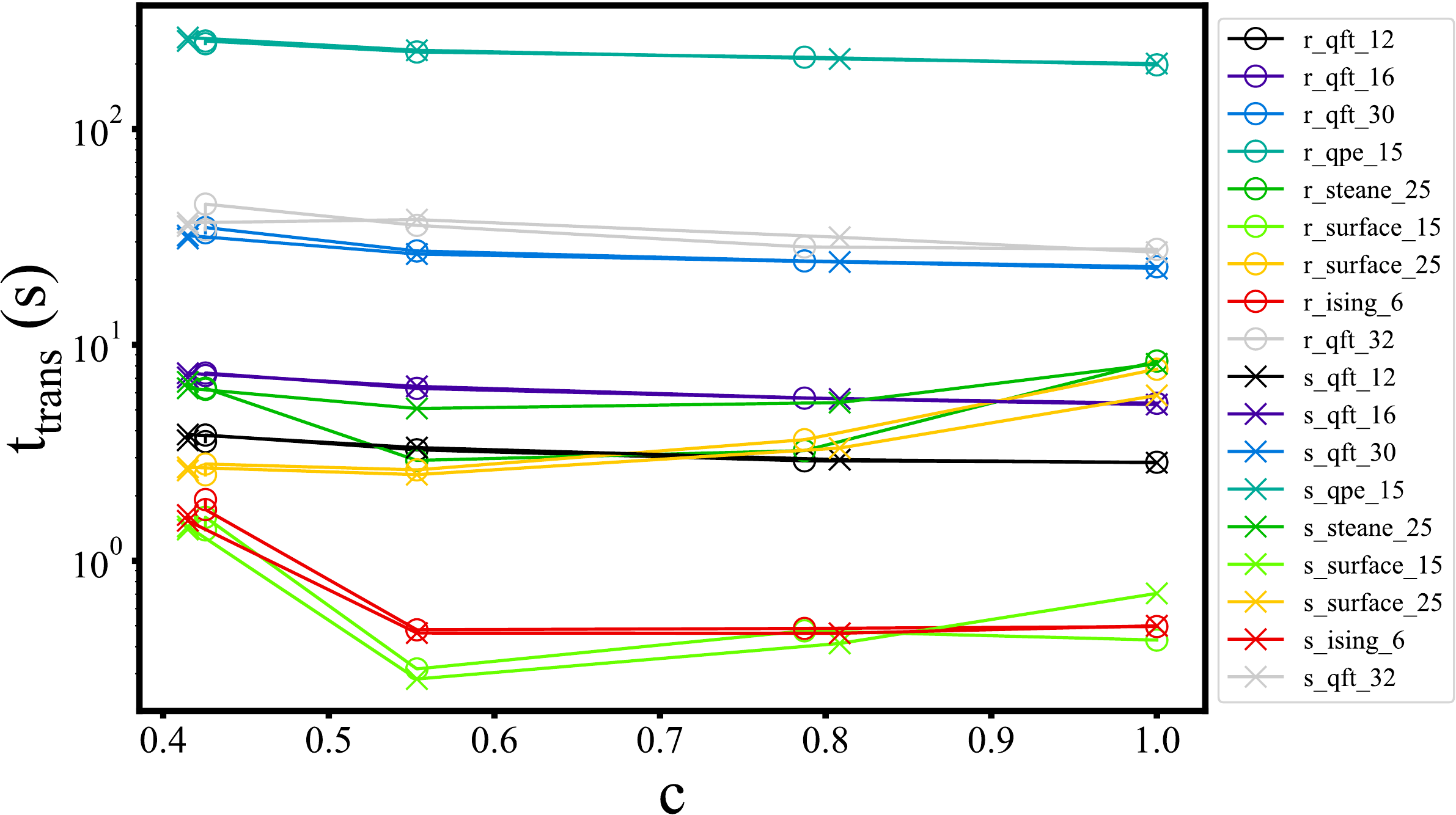}}
  \caption{
    (a) and (b) show the experiment results of how the simulation run time
    $\hat{t}_{sim}$, and the transpilation time $t_{trans}$ change with different
    connectivites $c$.  The time performance benchmark spans in multiple orders
    of magnitude in time so (a) and (b) are re-plotted in logarithmic scale as
    (c) and (d), respectively.
  }%
  \label{fig:data_logy_scale}
\end{figure*}

All of these riddles can be by answered by data shown in Fig.  \ref{fig:data_normal_scale} and \ref{fig:data_logy_scale}. Let us focus on the results in
Fig. \ref{fig:data_normal_scale} first.   We can see as the connectivity $c$
increases, $\hat{n}_{gate}$, $\hat{d}$, normalized score functions, and $t_{trans}$
decrease as a trend, namely the performance of the architectures with increasing
connectivities becomes better and better.  In Fig. \ref{fig:data_normal_scale}
(a), the data of algorithm $qft$ forms a group even they are different in the
aspects of numbers of qubit and topologies. All curves belonging to $qft$ family
are above $\hat{n}_{gate} = 5$.  The similar effect is also observed in Fig.
\ref{fig:data_normal_scale} (b), where $qft$ family is roughly above $\hat{d} =
4$. Since $\hat{n}_{gate}$ and $\hat{d} $ are related to the hardware
optimization, this observation hints that the performance of hardware
optimization is influenced by target algorithm family. A comparison between Fig.
\ref{fig:data_normal_scale} (a) and Fig. \ref{fig:data_normal_scale} (b) also
indicates that the sequence of the rest of algorithms does not change, from
$steane\_25$ (green), $ising\_6$ (red), $qpe\_15$ (green2), $surface\_25$
(yellow), to $surface\_15$ (green3).  This shows the correlation between $\hat{n}
_{gate}$ and $\hat{d}$. In other words, they are not independent factors in
circuit design, thus the score function of Eq. \ref{eq:s_simplified} is
introduced, which we used to select the output programs (quantum circuit), and
we believe is a more valuable indicator in circuit design.  In Fig. 
\ref{fig:data_normal_scale} (c), it is found that the normalized score function of each
number of qubits of $qft$ group drops monotonically as connectivity increases,
respectively. In addition, it is also shown that the normalized score function
drops monotonically as the number of qubits decrease, which is consistent with
intuition that as the number of qubits of the benchmarking algorithms is larger,
larger ratio of the chip is needed to run the benchmarking algorithms so the
connectivity and topology effect is more significant.     Fig. \ref{fig:data_normal_scale} (a) and Fig.  \ref{fig:data_normal_scale} (b) illustrate the
results in space domain of circuit design, while Fig. \ref{fig:data_normal_scale}
(d) shows the performance comparison in the time domain. 

When connectivity stays the same, it is not possible to draw the conclusion
which topology always dominates performance based on our data.  The possible
reasons of this phenomena could be as follows. First, the topologies of
rectangular-like and square-like are not so distinct in this work since the
rectangular-like layout that we chosen has aspect ratio $2$, which is not very
asymmetric and not quite different from square-like one. Second, due to the
limitation of computing resource and cost, we could not run all the benchmark
algorithms with large number of qubits. When not all qubits are used to run the
benchmarking algorithms, the topology effect could be reduced and less clear. In
addition, possible noisy data points, especially for simulations taking very
short time, may also add difficulty to draw the conclusion that which topology
is always better.  Need to emphasize, intuitively we expect some benchmarking
algorithms perform better on some specific architecture (certain layout/topology
\& connectivity), and some benchmarking algorithms perform better on other
specific architecture (certain layout/topology \& connectivity). However, we are
not making assumption that certain topology always dominates performance at
different connectivity even for different benchmarking algorithms. This is
consistent with our observation from the experiment data. 

In Fig. \ref{fig:data_logy_scale}, the effect of formation of group by algorithm
family does not appear. All algorithms are mixed up. The data is more noisy in Fig. 
\ref{fig:data_logy_scale} (a) than that in Fig. \ref{fig:data_logy_scale} (b) since some
simulation runs take very short of time which is more difficult to measure
accurately.  That would explain the bumpy data in Fig. \ref{fig:data_logy_scale}
(a) while that in Fig. \ref{fig:data_logy_scale} (b) is more coherent.  For most
of the algorithms, $\hat{t}_{sim}$ and $t_{trans} $ decreases as $c$ increases
and some of curves are quasi-linear. However, the $surface$ algorithm family does show a
non-monotonic behavior in Fig. \ref{fig:data_logy_scale} (c) and (d), which
means there is a possible optimal connectivity for certain specific algorithms.
In terms of topology comparison, there is no dominant effect  in general.  But
for some specific algorithms, such as $qft\_32$, $s$-architecture outperforms
$r$-architecture slightly.

In general, the impact of connectivity on performance for the same chip geometry
varies up to $50\%$ in our study by calculating and comparing the ratio of
minimium and maximium values of $\hat{n}_{gate}$, $\hat{d}$, $\hat{t}_{sim}$ and
$t_{trans}$ and the ratio can be as low as $3-5\%$. In practice, increasing
connectivity and enhancing gate fidelity are often contradictory goals, so a
tradeoff has to be made. Further research on noise and fidelity versus
connectivity is of interest but beyond the scope of this work.

\section{Conclusions and Future Work}

In this work, a preliminary set of quantum benchmark algorithms for
evaluating quantum chip performance is constructed, based on which the
performance of superconducting quantum chip structures is quantitatively
evaluated and these typical architectures are compared.

We find that connectivity plays an important role in the qubit chip design.  A
partially connected chip's performance is significantly lower than that of a
fully connected one, and the performance difference can be several times higher.
The experiment shows that quantum algorithms and circuits that use more
connectivity clearly benefit from a better-connected architecture. In addition,
the results suggest that co-designing quantum applications with the hardware
will be paramount in successfully using quantum computers in the future. This
work is mainly focused on superconducting-qubit-based quantum chips, but it can
be applied to other QC implementations as well.

The connectivity in this work is analyzed quantitatively. However, it is equally
crucial to quantify the topology. The topology $t$ describes the space a chip
would take, including symmetric, concentration factors etc, and the connectivity
$c$ describes fine structures in this space.  Intuitively, a function of $t$ and
$c$ may be useful to evaluate the quality of a processor, taking both of
topology and connectivity into consideration. For future work, it is worthy
drawing a phase diagram of performance of different benchmarking algorithms with
different topology indicator $t$ and connectivity indicator $c$. It would be very
beneficial if some relationship or empirical formula can be established from
such indicators and the quantitative parameters extracted from the benchmarking
algorithms. For a certain algorithm or a set of algorithms, how to design QC
chip with the optimal architecture and develop a corresponding optimized mapping
algorithm to execute the algorithm with the highest possible efficiency is a
topic worthy of further research.

The architectures studied are all two-dimensional lattice structures and belongs
to 2-D Nearest Neighbor structure which fits in the planar layout. In the future,
a three-dimensional chip architecture may become practical.  Arranging
connectivity in a three-dimensional architecture will be more challenging.  We
will further investigate how to build connectivity and make other tradeoffs in a
three-dimensional architecture.

Furthermore, quantitative study on the tradeoff between connectivity and qubit
quality, and supporting large-scale quantum programs on a network of quantum
chips making efficient use of connectivity would be interesting topics.


\noindent {\bf Acknowledgements}

The authors would like to thank Wenbing Fu, Xiaoyu Sun for useful discussions
and for help with code availability and testing.

\noindent {\bf Funding}

This work was supported in part by Pudong New Area Science and Technology
Development Fund (No. PKX2020-R17), the National Key Research and Development
Program of China (No. 2016YFA0301803), the Special Project for Research and
Development in Key areas of Guangdong Province (No. 2019B030330001), the Natural
Science Foundation of Guangdong Province (No. 2018A030313342) and the National
Natural Science Foundation of China (NSFC) (No. 61875060).

\section*{Declarations}

%
\noindent {\bf Conflict of interest}

The authors declare that they have no conflict of interest.

\noindent {\bf Data availability}

Data are generated using the code in via \url{www.github.com/sudoyang}.

\noindent {\bf Code availability}

The code for this project is available in the GitHub repository found in
\url{www.github.com/sudoyang}.



\bibliographystyle{unsrt2authabbrvpp}
\bibliography{turing.bib}

\begin{thebibliography}{10}

\bibitem{barends_superconducting_2014}
R.~Barends et~al.
\newblock Superconducting quantum circuits at the surface code threshold for
  fault tolerance.
\newblock {\em Nature}, 508(7497):500--503, April 2014.
\newblock Number: 7497 Publisher: Nature Publishing Group.

\bibitem{corcoles_demonstration_2015}
A.~D. Córcoles et~al.
\newblock Demonstration of a quantum error detection code using a square
  lattice of four superconducting qubits.
\newblock {\em Nature Communications}, 6(1):6979, April 2015.
\newblock Number: 1 Publisher: Nature Publishing Group.

\bibitem{riste_detecting_2015}
D.~Ristè et~al.
\newblock Detecting bit-flip errors in a logical qubit using stabilizer
  measurements.
\newblock {\em Nature Communications}, 6(1):6983, April 2015.
\newblock Number: 1 Publisher: Nature Publishing Group.

\bibitem{ofek_extending_2016}
N.~Ofek et~al.
\newblock Extending the lifetime of a quantum bit with error correction in
  superconducting circuits.
\newblock {\em Nature}, 536(7617):441--445, August 2016.
\newblock Number: 7617 Publisher: Nature Publishing Group.

\bibitem{takita_demonstration_2016}
M.~Takita et~al.
\newblock Demonstration of weight-four parity measurements in the surface code
  architecture.
\newblock {\em Physical Review Letters}, 117(21):210505, November 2016.
\newblock arXiv: 1605.01351.

\bibitem{debnath_demonstration_2016}
S.~Debnath et~al.
\newblock Demonstration of a small programmable quantum computer with atomic
  qubits.
\newblock {\em Nature}, 536(7614):63--66, August 2016.
\newblock Number: 7614 Publisher: Nature Publishing Group.

\bibitem{monz_realization_2016}
T.~Monz et~al.
\newblock Realization of a scalable {Shor} algorithm.
\newblock {\em Science}, 351(6277):1068--1070, March 2016.
\newblock Publisher: American Association for the Advancement of Science
  Section: Report.

\bibitem{jurcevic_demonstration_2021-1}
P.~Jurcevic et~al.
\newblock Demonstration of quantum volume 64 on a superconducting quantum
  computing system.
\newblock {\em Quantum Science and Technology}, 6(2):025020, March 2021.
\newblock Publisher: IOP Publishing.

\bibitem{honeywell_2020}
H.~Team.
\newblock The world’s highest performing quantum computer is here.
\newblock
  \url{https://www.honeywell.com/us/en/news/2020/06/the-worlds-highest-performing-quantum-computer-is-here}.
\newblock Accessed: 2021-05-12.

\bibitem{OSA_2020}
OSA.
\newblock Researchers on a path to build powerful and practical quantum
  computer.
\newblock
  \url{https://www.osa.org/en-us/about_osa/newsroom/news_releases/2020/researchers_on_a_path_to_build_powerful_and_practi/}.
\newblock Accessed: 2021-05-12.

\bibitem{collaborators_hartree-fock_2020}
G.~A.~Q. Collaborators*† et~al.
\newblock Hartree-{Fock} on a superconducting qubit quantum computer.
\newblock {\em Science}, 369(6507):1084--1089, August 2020.
\newblock Publisher: American Association for the Advancement of Science
  Section: Research Article.

\bibitem{cross_validating_2019}
A.~W. Cross et~al.
\newblock Validating quantum computers using randomized model circuits.
\newblock {\em Physical Review A}, 100(3):032328, September 2019.
\newblock Publisher: American Physical Society.

\bibitem{paul_quantum_2020}
P.~Smith-Goodson.
\newblock Quantum volume: A yardstick to measure the performance of quantum
  computers.
\newblock
  \url{https://www.forbes.com/sites/moorinsights/2019/11/23/quantum-volume-a-yardstick-to-measure-the-power-of-quantum-computers/?sh=1ca3c355bf4c}.
\newblock Accessed: 2021-05-12.

\bibitem{kjaergaard_superconducting_2020}
M.~Kjaergaard et~al.
\newblock Superconducting {Qubits}: {Current} {State} of {Play}.
\newblock {\em Annual Review of Condensed Matter Physics}, 11(1):369--395,
  March 2020.
\newblock arXiv: 1905.13641.

\bibitem{linke_experimental_2017}
N.~M. Linke et~al.
\newblock Experimental comparison of two quantum computing architectures.
\newblock {\em Proceedings of the National Academy of Sciences of the United
  States of America}, 114(13):3305--3310, March 2017.

\bibitem{siraichi_qubit_2018-1}
M.~Y. Siraichi et~al.
\newblock Qubit allocation.
\newblock In {\em Proceedings of the 2018 {International} {Symposium} on {Code}
  {Generation} and {Optimization}}, {CGO} 2018, pp. 113--125, New York, NY,
  USA, February 2018. Association for Computing Machinery.

\bibitem{zulehner_efficient_2018}
A.~Zulehner et~al.
\newblock Efficient mapping of quantum circuits to the {IBM} {QX}
  architectures.
\newblock In {\em 2018 {Design}, {Automation} {Test} in {Europe} {Conference}
  {Exhibition} ({DATE})}, pp. 1135--1138, March 2018.
\newblock ISSN: 1558-1101.

\bibitem{li_tackling_2019}
G.~Li et~al.
\newblock Tackling the {Qubit} {Mapping} {Problem} for {NISQ}-{Era} {Quantum}
  {Devices}.
\newblock In {\em Proceedings of the {Twenty}-{Fourth} {International}
  {Conference} on {Architectural} {Support} for {Programming} {Languages} and
  {Operating} {Systems}}, pp. 1001--1014, Providence RI USA, April 2019. ACM.

\bibitem{nielsen_quantum_2010-1}
M.~A. Nielsen and I.~L. Chuang.
\newblock {\em Quantum computation and quantum information}.
\newblock Cambridge University Press, Cambridge ; New York, 10th anniversary ed
  edition, 2010.
\newblock 00000.

\bibitem{lloyd_quantum_2013-1}
S.~Lloyd et~al.
\newblock Quantum algorithms for supervised and unsupervised machine learning.
\newblock {\em arXiv:1307.0411 [quant-ph]}, November 2013.
\newblock arXiv: 1307.0411.

\bibitem{stamatopoulos_option_2020-1}
N.~Stamatopoulos et~al.
\newblock Option {Pricing} using {Quantum} {Computers}.
\newblock {\em Quantum}, 4:291, July 2020.
\newblock Publisher: Verein zur Förderung des Open Access Publizierens in den
  Quantenwissenschaften.

\bibitem{ramos-calderer_quantum_2021-1}
S.~Ramos-Calderer et~al.
\newblock Quantum unary approach to option pricing.
\newblock {\em Physical Review A}, 103(3):032414, March 2021.
\newblock Publisher: American Physical Society.

\bibitem{christensen_steane-enlargement_2020-1}
R.~B. Christensen and O.~Geil.
\newblock On {Steane}-enlargement of quantum codes from {Cartesian} product
  point sets.
\newblock {\em Quantum Information Processing}, 19(7):192, May 2020.

\bibitem{fowler_surface_2012-1}
A.~G. Fowler et~al.
\newblock Surface codes: {Towards} practical large-scale quantum computation.
\newblock {\em Physical Review A}, 86(3):032324, September 2012.
\newblock Publisher: American Physical Society.

\bibitem{brush_history_1967-1}
S.~G. BRUSH.
\newblock History of the {Lenz}-{Ising} {Model}.
\newblock {\em Reviews of Modern Physics}, 39(4):883--893, October 1967.
\newblock Publisher: American Physical Society.

\bibitem{clarke_superconducting_2008-1}
J.~Clarke and F.~K. Wilhelm.
\newblock Superconducting quantum bits.
\newblock {\em Nature}, 453(7198):1031--1042, June 2008.

\bibitem{kielpinski_architecture_2002-1}
D.~Kielpinski et~al.
\newblock Architecture for a large-scale ion-trap quantum computer.
\newblock {\em Nature}, 417(6890):709--711, June 2002.
\newblock Number: 6890 Publisher: Nature Publishing Group.

\bibitem{cirac_quantum_1995-1}
J.~I. Cirac and P.~Zoller.
\newblock Quantum {Computations} with {Cold} {Trapped} {Ions}.
\newblock {\em Physical Review Letters}, 74(20):4091--4094, May 1995.
\newblock Publisher: American Physical Society.

\bibitem{imamoglu_quantum_1999-1}
A.~Imamog¯lu et~al.
\newblock Quantum {Information} {Processing} {Using} {Quantum} {Dot} {Spins}
  and {Cavity} {QED}.
\newblock {\em Physical Review Letters}, 83(20):4204--4207, November 1999.
\newblock Publisher: American Physical Society.

\bibitem{henriet_quantum_2020-1}
L.~Henriet et~al.
\newblock Quantum computing with neutral atoms.
\newblock {\em Quantum}, 4:327, September 2020.
\newblock Publisher: Verein zur Förderung des Open Access Publizierens in den
  Quantenwissenschaften.

\bibitem{igeta_quantum_1988-1}
K.~Igeta and Y.~Yamamoto.
\newblock Quantum mechanical computers with single atom and photon fields.
\newblock In {\em International {Conference} on {Quantum} {Electronics} (1988),
  paper {TuI4}}, pp. TuI4. Optical Society of America, July 1988.

\bibitem{arute_quantum_2019-1}
F.~Arute et~al.
\newblock Quantum supremacy using a programmable superconducting processor.
\newblock {\em Nature}, 574(7779):505--510, October 2019.
\newblock Number: 7779 Publisher: Nature Publishing Group.

\bibitem{qiskit-transpiler}
IBM.
\newblock Ibm qiskit transpiler.
\newblock \url{https://qiskit.org/documentation/apidoc/transpiler.html}.
\newblock Accessed: 2021-05-12.

\bibitem{barenco_elementary_1995-1}
A.~Barenco et~al.
\newblock Elementary gates for quantum computation.
\newblock {\em Physical Review A}, 52(5):3457--3467, November 1995.
\newblock Publisher: American Physical Society.

\end{thebibliography}


\end{document}